\documentclass[11pt]{article}

\usepackage{color,graphicx}
\usepackage{young}
\usepackage[vcentermath]{youngtab}
\usepackage{amsmath,amssymb,graphicx}
\usepackage{hyperref}
\definecolor{darkred}{rgb}{0.65,0.15,0}
\hypersetup{pdfborder={0 0 0},colorlinks=true,urlcolor=darkred,citecolor=blue,linkcolor=darkred,linktocpage=true}

\usepackage{cite}
\usepackage{amsmath}
\usepackage{amsfonts}
\usepackage{amssymb}
\usepackage{graphicx}%
\usepackage{amsthm}
\usepackage{mathrsfs}
\usepackage[T1]{fontenc}
\usepackage{enumerate}
\usepackage{color}

\def\4diml{four-dimensional}

\def\-1{^{-1}}

\newcommand{\A}{\mathscr{A}}

\newcommand{\M}{\mathscr{M}}
\newcommand{\D}{\mathscr{D}}
\newcommand{\G}{\mathscr{G}}

\makeatletter

\@addtoreset{equation}{section}
\makeatother

\begin{document}

\thispagestyle{empty}

\vspace{5mm}

\begin{center}
{\LARGE \bf $AdS_3 \times S^3$ Background from Poisson-Lie T-Duality}

\vspace{15mm}
\normalsize
{\large  Ali Eghbali}

\vspace{2mm}
{\small \em Department of Physics, Faculty of Basic Sciences,\\
Azarbaijan Shahid Madani University, 53714-161, Tabriz, Iran}\\
\vspace{4mm}
\verb"Email: eghbali978@gmail.com"\\

\vspace{6mm}

\begin{tabular}{p{12cm}}
{\small
We proceed to construct a dual pair for the $AdS_3 \times S^3$ background by
applying non-Abelian T-duality (here as Poisson-Lie (PL) T-duality on a semi-Abelian double).
By using a certain parametrization of the $4$-dimensional Lie group ${A}_2 \otimes 2{A}_1$ and
by a suitable choice of spectator-dependent matrices, the original $\sigma$-model including the $AdS_3 \times S^3$ metric
and a non-trivial $B$-field are constructed. The dual background constructed by means of the PL T-duality with the spectators
is an asymptotically flat one with a potential black hole interpretation supported by a non-trivial $H$-flux
whose metric contains the true singularity with
a single horizon.
The question of classical integrability of the non-Abelian T-dual $\sigma$-models under consideration is addressed, 
and their corresponding Lax pairs are found, depending on some
spectral parameters.
Finally, the conformal invariance conditions of the
models are checked up to two-loop order, and it has been concluded that the
resulting model is indeed a solution of supergravity.
}
\end{tabular}
\vspace{-1mm}
\end{center}

$~~~~~~~~~~~~~~~~~~~~~~~~~~~~$ string duality, $\sigma$-model

\setcounter{page}{1}
\newpage
\tableofcontents

 \vspace{5mm}


\section{Introduction}
\label{section1}
One of the most interesting examples in the context of AdS string backgrounds is
$AdS_3 \times S^3 \times {\cal M}^4$, where ${\cal M}^4$ is either $T^4$ or $S^3 \times S^1$,
such that $AdS_3 \times S^3 \times T^4$ and $AdS_3 \times S^3 \times S^3  \times S^1$ \cite{{Babichenko},{Zarembo1}}
are constructed from semi-symmetric space $\sigma$-models on
supercosets generated by the Lie supergroups $PSU(1 , 1|2)$ and $D(2 , 1; \alpha)$, respectively.
It has shown that the classical $\kappa$-symmetric type IIB string action on $AdS_3 \times S^3$ can be derived from
the Lie superalgebra $su(1 , 1|2)^2$ \cite{rahmfeld}.
From integrability perspective, more progress has been made in understanding
$AdS_3/CFT_2$ duality \cite{Babichenko} (see, also, \cite{Sfondrini} and references therein).
For instance,  the Green-Schwarz action of type IIB strings with Ramond-Ramond three-form flux
compactified on $AdS_3 \times S^3 \times {\cal M}^4$ is an integrable classical theory \cite{Babichenko}.
An interesting feature of the type IIB string theories on
$AdS_3 \times S^3 \times {\cal M}^4$ is that the backgrounds can be supported by a combination
of Ramond-Ramond and Neveu-Schwarz-Neveu-Schwarz fluxes, in a way that
this provides us with a family of string backgrounds \cite{{Cagnazzo},{Hoare}}.
In the case of the  $AdS_3 \times S^3 \times T^4$, the action for the type IIB superstring propagating in
this background has been given in \cite{pesando}. There, the super gravitational background corresponding
to a spontaneous compactification has been computed.
In addition, it has shown that by investigating the type IIA  superstring on the
$AdS_3 \times S^3 \times {\cal M}^4 $ background with $ {\cal M}^4=S^3  \times S^1$ or ${\cal M}^4 =T^4$
one can derive the $\kappa$-symmetry gauge-fixed Green-Schwarz string action to quadratic order in fermions and quartic order in fields
\cite{linus1} (see, also, \cite{linus2}). In Ref. \cite{Rafael}, it has shown that
closed spinning strings in $AdS_3 \times S^3 \times T^4$ in the presence of Neveu-Schwarz-Neveu-Schwarz three-form flux can
be described by an extension of the Neumann-Rosochatius system.
It has recently been shown that a family of type IIB string backgrounds that are deformations of
$AdS_3 \times S^3 \times T^4$ background
with a squashed $AdS_3 \times S^3$ metric, can be supported by a combination of Neveu-Schwarz-Neveu-Schwarz and Ramond-Ramond fluxes \cite{ben.hore2}.
It is shown there that the type IIB supergravity backgrounds can be obtained, up to T-dualities, from an integrable inhomogeneous Yang-Baxter deformation
of the original $AdS_3 \times S^3$ supercoset model.
Of course, the compact 4-dimensional manifold ${\cal M}^4$ does
not play a role in our story; in fact, the solutions that we will study will have no dynamics along
the ${\cal M}^4$ and thus we will not include these directions in what follows.
We particularly focus on finding the non-Abelian target space dual of the $AdS_3 \times S^3$ background.

One of the most interesting properties of string theory is that different spacetime geometries can correspond to equivalent classical solutions.
String theory has the T-duality symmetry when the target space has Abelian isometries.
Abelian T-duality \cite{Buscher} requires the two target spaces it links to
have Abelian isometries. But there is also a non-Abelian counterpart, which is known as the
non-Abelian T-duality \cite{nonabelian}, where the isometry group is non-Abelian,
which works well as a solution-generating technique in supergravity.
It only holds at the classical level but still admits to construct new string backgrounds.
On the other hand, it has been shown that double field theory \cite{{hull.zwiebach},{Hohm}}
is an approach to reveal Abelian T-duality by formally doubling the complete target space.
Later, it has been provided a duality transformation
rule for the Ramond-Ramond fields by using the technique of double field theory
in a way that through a formal T-duality in double field theory, the solution of the generalized supergravity equations of motion was mapped to a solution of the conventional supergravity \cite{Sakatani}.
Accordingly, by applying the traditional non-Abelian T-duality for non-unimodular algebras, it has been obtained
several non-Abelian T-dual backgrounds for the $AdS_3 \times S^3 \times T^4$ in both cases of the
absence and presence of the Ramond-Ramond fields \cite{Sakatani}.  One can also see in \cite{Lozano4} a class of
new examples of non-Abelian T-dual backgrounds by considering target spaces containing
coset manifolds.
Additionally, by examining the supersymmetry of the massive IIA T-dual theory in relation
to that of the original type-IIB theory,
it has been shown that \cite{Lozano3} the mapping of the Killing spinor
equations requires an additional condition breaking the original supersymmetry by a half for spinors transforming under $SO(4)$.
In Ref. \cite{Lozano1}, it has been discussed some aspects of the class of $AdS_3 \times S^2$ solutions with small
${\cal N} = (0, 4)$ supersymmetry and $SU(2)$-structure constructed in \cite{11}.
Through this study, the non-Abelian T-dual space of the $AdS_3 \times S^3 \times CY_2$ geometry, constructed in \cite{Thompson5}
(see, also, \cite{Lozano2}), has been rediscovered as the leading order in an expansion on the number of gauge groups.

A generalization of Abelian and traditional non-Abelian dualities was proposed by Klimcik and Severa \cite{{Klim1},{Klim2}}.
This kind of T-duality called PL T-duality.
It deals with $\sigma$-models based on two Lie groups which form a Drinfeld
double \cite{Drinfeld} and the duality transformation exchanges their roles.
In contrast to its Abelian/non-Abelian descendants, it does not require
isometries.
In recent years, we have witnessed further interest in PL T-duality, driven by $\sigma$-models based on Lie groups
\cite{{Alekseev1},{Tyurin},{Sfetsos1},{Lledo},{JR},{Majid},{N.Mohammedi2},{Hlavaty20002},{eghbali11},{EMR13},{Eghbali.2},{ERA},{sakatani2}},
as well as Lie supergroups \cite{{ER.super1},{EHR.super2},{eghbali}}.
It's worth mentioning that a formulation of double field theory with a Drinfeld double has been presented in \cite{hassler},
in such a way that it makes PL T-duality manifest.
Also in \cite{lust6}, it has been given some comment on generalizations of results and techniques known from the Abelian T-duality,
in such a way that it includes the remarks on a double field theory based on the non-Abelian T-duality.

The main purpose of the present paper is to study the non-Abelian T-dualization (here as the PL T-duality on a semi-Abelian double) of
the $AdS_3 \times S^3$ background.
The background metric and $B$-field associated to the $AdS_3 \times S^3$ are defined as follows
\begin{eqnarray}
ds^2 &=& k\Big[d {\bar \phi}^2 + e^{2{\bar \phi}}~ dx^+  dx^- +\frac{1}{4} \Big(d\theta^2+ \sin^2\theta d\varphi^2+
(d\psi + \cos\theta d\varphi)^2\Big)\Big],\label{1.1}\\
B &=& k x^+  e^{2{\bar \phi}}  d{\bar \phi} \wedge  dx^- +\frac{k}{4}  \cos\theta  d\psi \wedge d\varphi,\label{1.2}
\end{eqnarray}
where $({\bar \phi}, x^+, x^-)$ denote the Poincar\'{e} coordinates of $AdS_3$, $(\psi, \varphi, \theta)$
parametrize the sphere $S^3$ as a Hopf fibration, and $k={l_{_{AdS}}^2}/{l_{s}^2}$, where $l_{_{AdS}}^2$ and ${l_{s}^2}$ denote the AdS and string
scales, respectively. The metric \eqref{1.1} is flat in the sense that its scalar curvature vanishes.
The $AdS_3 \times S^3$ background along with a constant dilaton field and zero
cosmological constant make up a solution for the
vanishing of the one-loop beta-function equations. We will show that this background can be also conformally invariant up to two-loop order.
According to equations \eqref{1.1} and \eqref{1.2}, the bosonic theory on $AdS_3 \times S^3$ amounts to a Wess-Zumino-Witten
model with gauge group $SL(2) \otimes SU(2)$.
Notice that the $AdS_3 \times S^3$ background in the absence of $B$-field corresponds to the case of pure Ramond-Ramond flux, where
the theory can be formulated in terms of a Green-Schwarz coset, while in the presence of $B$-field \eqref{1.2} is the
limit of pure Neveu-Schwarz-Neveu-Schwarz flux, and can be described by a supersymmetric Wess-Zumino-Witten model.
In the absence of flux the $\sigma$-model for closed strings rotating in $AdS_3 \times S^3$ becomes the
Neumann-Rosochatius integrable system describing an oscillator on a sphere or an
hyperboloid with a centrifugal potential.

In the context of the PL T-duality with spectators, the choice of spectator-dependent matrices plays a key role in the process of constructing models.
By a suitable choice, we obtain the original $\sigma$-model background including the $AdS_3 \times S^3$ metric \eqref{1.1} and
$B$-field \eqref{1.2}. Then, to get the dual solution we employ the PL T-duality transformations.
This is realized by describing the original 6-dimensional geometry and the corresponding dual one by
a semi-Abelian Drinfeld double, plus some spectator fields.
As we will see, our dual solution will be different from those of \cite{Sakatani}.
Also, we employ the duality transformation of the dilaton fields that makes the T-dual $\sigma$-models conformal up to the one-loop order.
Nevertheless, the vanishing of the one-loop beta-functions for both models is imposed, to guarantee UV finiteness at
quantum level.
Finally, we are interested in testing the conformal invariance of both original and dual solutions up to two-loop order.
Notice that in the case of the $AdS_3 \times S^3$ background one may apply the usual rules of non-Abelian
T-duality without further corrections, and still be able to obtain two-loop
solutions. However, in general, further corrections to the rules are
necessary. Borsato and Wulff have recently shown that the PL T-duality can be extended to order
$\alpha'$, namely, two loops in the $\sigma$-model perturbation theory,
provided that the map is corrected \cite{Wulff22} (see, also, \cite{{Hassler2},{Marques}}).

In Ref. \cite{N.Mohammedi}, by introducing the general framework for a zero curvature representation
of the equations of motion of a 2-dimensional
 non-linear $\sigma$-model, it has been derived a target space condition for this requirement.
To this condition, we study the integrability of non-Abelian T-dual $\sigma$-models under consideration,
and show that the models admit a Lax pair
representation. Indeed, these models are new integrable 2-dimensional theories.

The paper is organized as follows. After Introduction section,  section \ref{section2} reviews the construction
of PL T-dual $\sigma$-models in the presence of spectator fields, where necessary formulas are summarized.
Section \ref{section3} contains the original results of the work:
using the formulation introduced in section \ref{section2} and
then starting from the semi-Abelian double $({\A}_2 \oplus 2{\A}_1 , 4{\A}_1)$, we construct
a pair of PL T-dual $\sigma$-models including the $AdS_3 \times S^3$ background and its dual pair.
Investigation of the structure and asymptotic nature of the dual geometry
including the horizon and singularity are  discussed  at the end of section \ref{section3}.
In section \ref{section4}, we investigate in detail the integrability of the T-dual $\sigma$-models built in section \ref{section3},
in such a way that we show that corresponding Lax pairs depend on some spectral
parameters.
We start section \ref{section5} by introducing the vanishing of the beta-function equations up to two-loop order, and then
study the conformality of the T-dual $\sigma$-models up to two-loop order.
We conclude with a final discussion of the results with remarks and perspectives.


\section{A brief review of PL T-duality with spectator fields}
\label{section2}
Before proceeding to review the construction of PL T-dual $\sigma$-models in the presence of spectator fields \cite{{Klim1},{Klim2}}, let us
introduce the action of 2-dimensional non-linear $\sigma$-model.
A classical solution to the equations of motion of string theory is equivalent to a conformally-invariant, 2-dimensional field theory.
In the case that the 2-dimensional field theory is a $\sigma$-model, whose 2-dimensional scalar fields, $X^{^M}(\tau , \sigma)$,
can be interpreted as describing the worldsheet of a string propagating
within a {\it target} spacetime ($d$-dimensional manifold $\M$).
Suppose that the action of $\sigma$-model related to the bosonic string theory has the following form
\begin{eqnarray}\label{a.1}
S &=& \frac{1}{4 \pi \alpha'}\int\!d\tau  d\sigma \sqrt{-h} \big[h^{\alpha \beta}G_{_{MN}}(X)
+\epsilon^{\alpha \beta} B_{_{MN}} (X)\big]\partial_{\alpha}X{^{M}}
\partial_{_\beta}X^{{N}} \nonumber\\
&&~~~~~~~~~~~~~~~~~~~~~~~~~~~~~~~~~~~~~~~~~~~~~~~~~~~~+ \frac{1}{8 \pi} \int\!d\tau  d\sigma  ~R^{^{(2)}} ~\Phi(X),
\end{eqnarray}
where $\sigma^{\alpha}=(\tau , \sigma)$ are the worldsheet coordinates, $h_{\alpha \beta}$ and $\epsilon_{\alpha \beta}$
\footnote{The inverse of $h_{\alpha \beta}$  and $\epsilon_{\alpha \beta}$ will be denoted by $h^{\alpha \beta}$ and $\epsilon^{\alpha \beta}$, respectively.} are
the 2-dimensional metric and anti-symmetric Levi-Civita symbol, respectively,  $R^{^{(2)}}$ is the curvature scalar for the metric $h_{\alpha \beta}$ and $h=\det h_{\alpha \beta}$.
The dimensionful coupling constant $\alpha'$ turns out to be the inverse string tension.
The coupling functions, $G_{_{MN}}(X)$, $B_{_{MN}}(X)$ and $\Phi(X)$, are interpreted in
string theory as being background values for three fields-the {\it metric},
the {\it axion} ($B$-field), and the {\it dilaton}- which represent three of the modes
of the string. The line element and $B$-field corresponding to
action \eqref{a.1} are, in the coordinate basis, defined in the following way
\begin{eqnarray}
ds^2 = G_{_{MN}} dX^{^M} dX^{^N},~~~~~~~B = \frac{1}{2} B_{_{MN}} ~ dX^{^M}\wedge dX^{^N}.\label{4.1.3}
\end{eqnarray}
Let us now consider a 2-dimensional non-linear $\sigma$-model for the $d$ field variables
$X^{^{M}} = (x^{\mu} , y^i)$, where $x^\mu,~ \mu = 1, . . . , dim  ~G$ stand for the coordinates of Lie group $G$ acting freely from right on
the target manifold $\M \approx {\cal O} \times G$, and $y^i,~i = 1, \cdots , d-dim ~G$ are
the coordinates labeling the orbit ${\cal O}$ of $ G$ in  ${\M}$. Here we work in
the standard light-cone variables on the worldsheet,
$\sigma^{\pm} =(\tau \pm \sigma)/2$ together with $\partial_{_\pm}=\partial_{\tau} \pm \partial_{\sigma}$.
It should be remarked that the coordinates $y^i$ do not participate in the PL T-duality transformations
and are therefore called spectators \cite{Sfetsos1}.  The corresponding action has the form\footnote{In the absence of the dilaton field, one may compare
the action \eqref{c.1} with \eqref{a.1} to obtain the metric and $B$-field corresponding to \eqref{c.1}.
}
\begin{eqnarray}\label{c.1}
S = \frac{1}{2} \int d\sigma^{+}  d\sigma^{-} \Big[{{E}_{_{ab}}}~
R_{+}^a \;R_{-}^b + \phi^{{(1)}}_{a j} R_{+}^a \partial_{-} y^{j}+
\phi^{{(2)}}_{i b}  \partial_{+} y^{i} R_{-}^b  +\phi_{_{ij}}
\partial_{+} y^{i} \partial_{-} y^{j} \Big],
\end{eqnarray}
where $R_{\pm}^a$ are the components of the right-invariant Maurer-Cartan one-forms which are constructed by means of
an element $g$ of the Lie group ${G}$ as
\begin{eqnarray}\label{c.1.1}
R_{\pm} = R_{\pm}^a~ T_a = (\partial_{_\pm} g~g^{-1})^a ~ T_a = \partial_{_\pm} x^{\mu}~ R_{\mu}^{~a} ~ T_a,
\end{eqnarray}
in which ${T}_a$, $a=1, \cdots, dim ~G$ are the bases of the Lie algebra ${\G}$ of ${G}$, and
for notational convenience we will also use $R_{\pm}^i =\partial_{_\pm} y^{i}$.
The couplings ${{E}_{_{ab}}}, \phi^{{(1)}}_{a j}, \phi^{{(2)}}_{i b} $ and $\phi_{_{ij}}$ may depend on all variables $x^\mu$ and $y^i$.

Similarly we introduce another $\sigma$-model for the $d$ field variables ${\tilde X}^{^{M}} =({\tilde x}^{\mu} , y^i)$, where
${\tilde x}^{\mu},~ \mu = 1, \cdots , dim  ~{\tilde G}$ parametrize an element ${\tilde g}$ of a Lie group ${\tilde G}$ whose dimension is, however,
equal to that of $G$, and the rest of the variables are the same $y^i$'s used in \eqref{c.1}. Accordingly, we
introduce a different set of bases ${{\tilde T}^a }$ of the Lie algebra ${\tilde \G}$, with $a = 1, \cdots , dim ~G$.
We furthermore consider the components of the right-invariant Maurer-Cartan one-forms on
${\tilde G}$ as $(\partial_{_\pm} \tilde g~\tilde g^{-1})_a={\tilde R}_{{\pm}_a}=\partial_{_\pm} {\tilde x}^{\mu} {\tilde R}_{\mu a}$.
In this case, the corresponding action has the form
\begin{eqnarray}\label{c.2}
\tilde S = \frac{1}{2} \int d\sigma^{+}  d\sigma^{-}\Big[{{{\tilde E}}^{{ab}}}~
{\tilde R}_{+_{a}}{\tilde R}_{-_{b}}+{\tilde \phi}^{\hspace{0mm}{(1)^{ a}}}_{~~~j} ~{\tilde R}_{+_{a}}\partial_{-} y^{j}+
{\tilde \phi}^{\hspace{0mm}{(2)^{ b}}}_{i} \partial_{+} y^{i} ~{\tilde R}_{-_{b}}
+{\tilde \phi}_{_{ij}} \partial_{+} y^{i} \partial_{-} y^{j}\Big].
\end{eqnarray}
Notice that here one does not require any isometry associated with the Lie groups $G$ and $\tilde G$.
The $\sigma$-models \eqref{c.1} and \eqref{c.2} will be dual to each other in the sense of PL
T-duality \cite{{Klim1},{Klim2}} if the associated Lie algebras $\G$ and ${\tilde \G}$ form a pair of maximally isotropic
subalgebras into which the Lie algebra $\D$ of a Lie group $D$ known as the {\it Drinfeld double} \cite{Drinfeld}
can be decomposed. This implies that besides the commutator relations
\begin{eqnarray}\label{c.3}
[T_a , T_b] = {f^c}_{ab} ~T_c,~~~~~~~
[{\tilde T}^a , {\tilde T}^b] = {{\tilde f}^{ab}}_{\; \; \: c} ~{\tilde T}^c,
\end{eqnarray}
for the Lie algebras  $\G$ and ${\tilde \G}$, respectively, we must also consider
\begin{eqnarray}\label{c.4}
{[T_a , {\tilde T}^b]} = {{{\tilde f}^{bc}}_{\; \; \; \:a} {T}_c + {f^b}_{ca} ~{\tilde T}^c}.
\end{eqnarray}
The Jacobi identity on $\D$ relates the structure constants of the two Lie
algebras as \cite{{Klim1},{Klim2}}
\begin{eqnarray}\label{c.5}
{f^a}_{bc}{\tilde{f}^{de}}_{\; \; \; \; a}=
{f^d}_{ac}{\tilde{f}^{ae}}_{\; \; \; \;  b} +
{f^e}_{ba}{\tilde{f}^{da}}_{\; \; \; \;  c}+
{f^d}_{ba}{\tilde{f}^{ae}}_{\; \; \; \; c}+
{f^e}_{ac}{\tilde{f}^{da}}_{\; \; \; \; b}.
\end{eqnarray}
In addition to these, there is a bilinear invariant $<.~ ,~ .>$  with
the various generators obeying
\begin{eqnarray}\label{c.6}
<T_a , {\tilde T}^b> &=& {\delta _a}^{~b},\nonumber\\
<T_a , T_b> &=& <{\tilde T}^a, {\tilde T}^b> ~ =~ 0.
\end{eqnarray}
There remains to relate the couplings ${{E}_{_{ab}}}, {{\phi}}^{{(1)}}_{a j}, \phi^{{(2)}}_{i b} $
and ${{\bf \phi}}_{_{ij}}$ in \eqref{c.1} to ${{{\tilde E}}^{{ab}}}, {\tilde \phi}^{\hspace{0mm}{(1)^{ a}}}_{~~~j},
{\tilde \phi}^{\hspace{0mm}{(2)^{ b}}}_{i}$ and ${\tilde \phi}_{_{ij}}$ in \eqref{c.2}.
It has been shown that \cite{{Klim1},{Klim2},{Tyurin},{Sfetsos1}} the various couplings in the
$\sigma$-model action \eqref{c.1} are restricted to be
\begin{eqnarray}\label{c.7}
{{E}} &=& \big(E^{{-1}}_{0}+ \Pi\big)^{-1},~~~~~~~~~~~
\phi^{{(1)}} = {{E}}~E^{{-1}}_{0}~F^{^{(1)}},\nonumber\\
\phi^{{(2)}}&=& F^{^{(2)}}~ E^{{-1}}_{0}~{{E}},~~~~~~~~~~~~~~~~
{ \phi} = F -F^{^{(2)}}~\Pi~{{E}}~E^{{-1}}_{0}~F^{^{(1)}},
\end{eqnarray}
where the new couplings $E_{0}, F^{^{(1)}}, F^{^{(2)}}$ and $F$ may be at most functions of
the spectator variables $y^{i}$ only. In equation \eqref{c.7},
$\Pi(g)$ defined by $\Pi^{^{ab}}(g) = b^{^{ac}}(g)~ (a^{-1})_{_{c}}^{^{~b}}(g)$ is called the Poisson structure on $G$ so that submatrices
$a(g)$ and $b(g)$ are calculated in the following way
\begin{eqnarray}\label{c.8}
g^{-1} T_{{_a}}~ g &=& a_{_{a}}^{^{~b}}(g) ~ T_{{_b}},\nonumber\\
g^{-1} {\tilde T}^{{^a}} g &=& b^{^{ab}}(g)~ T_{{_b}}+(a^{-1})_{_{b}}^{^{~a}}(g)~{\tilde T}^{{^b}}.
\end{eqnarray}
Finally, the relation of the dual action couplings to those of the original one is given by \cite{{Klim1},{Klim2},{Tyurin},{Sfetsos1}}
\begin{eqnarray}\label{c.9}
{{\tilde E}} &=& \big(E_{0}+ {\tilde \Pi}\big)^{-1},~~~~~~~~~~
{\tilde \phi}^{{(1)}} =  {{\tilde E}}~F^{^{(1)}},\nonumber\\
{\tilde \phi}^{{(2)}} &=& - F^{^{(2)}} ~{{\tilde E}},~~~~~~~~~~~~~~~~~
{\tilde \phi}           = F-F^{^{(2)}} ~{{\tilde E}}~F^{^{(1)}}.
\end{eqnarray}
Analogously one can define matrices ${\tilde a} (\tilde g), {\tilde b} (\tilde g)$ and ${\tilde \Pi} (\tilde g)$ by just replacing the untilded symbols
by tilded ones.

As mentioned in the above,  the actions \eqref{c.1} and \eqref{c.2} correspond to PL T-dual $\sigma$-models.
In the non-Abelian duality case, where ${f^a}_{bc} \neq 0$ and $\tilde f^{ab}_{~~~c}=0$,
it follows from \eqref{c.4} and \eqref{c.8} that $ b(g)=0 $, then, $ \varPi(g)=0 $; consequently, $ E=E_0$, and thus the action \eqref{c.1} reduces to
\begin{eqnarray}\label{c.11}
S = \frac{1}{2} \int d\sigma^{+}  d\sigma^{-} \Big[{{E_0}_{_{ab}}}~
R_{+}^a \;R_{-}^b + F^{{(1)}}_{a j} R_{+}^a \partial_{-} y^{j}+
F^{{(2)}}_{i b}  \partial_{+} y^{i} R_{-}^b  +F_{_{ij}}
\partial_{+} y^{i} \partial_{-} y^{j} \Big].
\end{eqnarray}
In this case, if the couplings $E_{0}, F^{^{(1)}}, F^{^{(2)}}$ and $F$ are chosen to be symmetric, then one concludes that the $B$-field  vanishes.
In general, these couplings can have an anti-symmetric part, and in that case the $B$-field would be non-vanishing.
In the following, we will apply the above formulae in order to
construct the non-Abelian T-dual space of the $AdS_3 \times S^3$ background.

Before closing this section, let us turn our attention to the dilaton shifts in both original and dual $\sigma$-models.
As shown in \cite{vonUnge}, the duality transformation must be supplemented by a
correction that comes from integrating out the fields on the dual group
in path integral formulation in such a way that it can be absorbed at the one-loop level into the transformation of the dilaton field.
Based on a regularization of a functional
determinant in a path integral formulation of PL T-duality by
incorporating spectator fields, the correct formula of dilaton transformation is  given by \cite{vonUnge}
\begin{eqnarray}
\Phi &=&\varphi^{^{(0)}} +\log |\det {{E}}| - \log |\det {E_0}|  - \log |\det {a(g)}|,\label{c.16.1}\\
{\tilde \Phi} &=& \varphi^{^{(0)}}+\log |\det {{\tilde E}}| - \log |\det {{\tilde a} ({\tilde g})}|,\label{c.16.2}
\end{eqnarray}
in which $\varphi^{^{(0)}}$ is the dilaton that makes the original $\sigma$-model conformal (up to the one-loop order)
and may depend on both group and spectator coordinates. Accordingly,
the dual background can also be conformal at the one-loop level with a new dilaton field obtaining from
equation \eqref{c.16.2}.


\section{Non-Abelian target space dual of the $AdS_3 \times S^3$ background: Starting from the semi-Abelian double $({\A}_2 \oplus 2{\A}_1 , 4{\A}_1)$}
\label{section3}

In this section we explicitly construct a pair of PL T-dual $\sigma$-models (here as
PL T-duality on semi-Abelian doubles) on the $4+2$-dimensional
manifolds ${\M} \approx {\cal O} \times G$ and $\tilde {\M} \approx {\cal O} \times {\tilde G}$ as the target spaces. Here
$G$ is considered to be the 4-dimensional Lie group ${A}_2 \otimes 2{A}_1$ acting freely on $\M$, while ${\tilde G}$  is
the 4-dimensional Abelian Lie group $4{A}_1$.
As we will show, the original model describes the $AdS_3 \times S^3$ background.
In the dual model, we will encounter a true singularity, and determine
the structure and asymptotic nature of the dual space.

\subsection{The original $\sigma$-model:  The $AdS_3 \times S^3$ background}

As mentioned above, we shall obtain the $AdS_3 \times S^3$ background from a T-dualizable $\sigma$-model
constructing on a $4+2$-dimensional manifold $\M \approx {\cal O} \times G$, in which $G$  is the
4-dimensional Lie group ${A}_2 \otimes 2{A}_1$ acting freely on $\M$.
Then, as we will show, in order to study the non-Abelian T-duality of the model, the dual manifold is considered to be
${\tilde \M} \approx {\cal O} \times  {\tilde G}$, in which ${\tilde G}$  is the
4-dimensional Abelian Lie group $4{A}_1$.
A copy of the commutation relations of the decomposable Lie algebra
${\A}_2 \oplus 2{\A}_1$ \cite{patera} is given by
\begin{eqnarray}\label{4.1}
[T_{_1} , T_{_2}] = T_2,~~~~[T_{_3}~ , ~.]~=~0,~~[T_{_4} ~, ~.]=0.
\end{eqnarray}
As we saw in the previous section,  PL T-dual $\sigma$-models are defined by PL group manifolds which constitute a Drinfeld double.
Accordingly, utilizing \eqref{c.3}, \eqref{c.4} together with \eqref{4.1}, and also using the fact that the dual Lie algebra, $4{\A}_1$, is Abelian,
one can constitute the 8-dimensional Lie algebra of the Drinfeld double $({\A}_2 \oplus 2{\A}_1 , 4{\A}_1)$.
This semi-Abelian double
is generated by the generators $(T_{_a} , {\tilde T}^a)$ with the following non-zero Lie brackets
\begin{eqnarray}\label{4.2}
[T_1 , T_2] = T_2,~~~~~[T_1 ,{\tilde T}^2]=-{\tilde T}^2,~~~~~
{[T_2 ~, ~{\tilde T}^2]} = {\tilde T}^1.
\end{eqnarray}
In order to write the action \eqref{c.1}
on the $4+2$-dimensional manifold ${\M}$ explicitly we need to find the components of the right-invariant Maurer-Cartan
forms $R_{\pm}^a$  on the Lie group ${A}_2 \otimes 2{A}_1$.
To this purpose we use the following parametrization of the group manifold:
\begin{eqnarray}\label{4.3}
g~=~e^{x_{_1} T_1}~e^{x_{_2} T_2}~e^{x_{_3} T_3}~e^{x_{_4} T_4},
\end{eqnarray}
where $(x_{_1}, \cdots, x_{_4})$ stand for the coordinates
of the ${A}_2 \otimes 2{A}_1$. Using \eqref{c.1.1} and \eqref{4.1} one then gets
\begin{eqnarray}\label{4.4}
R_{\pm}^1&=& \partial_{\pm} x_{_1},~~~~~~~~R_{\pm}^2~=~ e^{x_{_1}}~\partial_{\pm} x_{_2},\nonumber\\
{R_{\pm}^3}&=& \partial_{\pm} {x}_{_3},~~~~~~~~R_{\pm}^4= \partial_{\pm} { x}_{_4}.
\end{eqnarray}
Since the dual Lie group is considered to be Abelian, it follows from \eqref{c.4} and \eqref{4.3} together with \eqref{c.8} that
$\Pi(g) =0$.
To achieve a $\sigma$-model with the $AdS_3 \times S^3$ background including the metric \eqref{1.1}
and $B$-field \eqref{1.2} one has to choose the spectator-dependent matrices in the following form
\begin{eqnarray}\label{4.5}
{{E_0}_{_{ab}}}&=&\left( \begin{array}{cccc}
                    0 & \frac{k}{2} e^{2 y_{_0}} & 0 & 0\\
                     \frac{k}{2} e^{2 y_{_0}} & 0 & 0 & 0\\
                      0 & 0 & \frac{k}{4} & \frac{k}{2} \cos y_{_1}\\
                    0 & 0 & 0 & \frac{k}{4}
                      \end{array} \right),~~~~
F^{^{(1)}}_{a j}=\left( \begin{array}{cc}
                    0 ~& 0\\
                    -k e^{2 y_{_0}} & 0\\
                    0 & 0\\
                    0 & 0
                      \end{array} \right),\nonumber\\
F^{^{(2)}}_{i b}&=&\left( \begin{array}{cccc}
                      0 & k e^{2 y_{_0}} & 0 & 0\\
                      0 & 0 & 0 & 0
                      \end{array} \right),~~~~~~~~~~~~~~~~~~~~F_{ij}=\left( \begin{array}{cc}
                    k & 0\\
                    0 & \frac{k}{4}
                      \end{array} \right).
\end{eqnarray}
where $y^{i}=(y_{_0} , y_{_1})$ are the coordinates of the orbit $\cal O$ (spectator fields) of $G$ in manifold ${\M}$.
Note that the parameter $k$ has already been introduced in equation \eqref{1.1}.
Inserting \eqref{4.4} and \eqref{4.5} into \eqref{c.11},
the original $\sigma$-model is worked out to be
\begin{eqnarray}\label{4.6}
S &=& \frac{k}{2} \int d \sigma^+ d \sigma^-~\Big[\partial_+ y_{_0} \partial_- y_{_0}+ \frac{1}{4}  (\partial_+ y_{_1} \partial_- y_{_1}
+ \partial_+ x_{_3} \partial_- x_{_3}+ \partial_+ x_{_4} \partial_- x_{_4}) \nonumber\\
&&~~~~~~~~~~~~~~~~~~+ \frac{1}{2} e^{x_{_1} + 2 y_{_0}} (\partial_+ x_{_1} \partial_- x_{_2} +\partial_+ x_{_2} \partial_- x_{_1}) +
\frac{1}{2} \cos y_{_1} ~\partial_+ x_{_3} \partial_- x_{_4}\nonumber\\
&&~~~~~~~~~~~~~~~~~~+e^{x_{_1} + 2 y_{_0}} (\partial_+ y_{_0} \partial_- x_{_2} -\partial_+ x_{_2} \partial_- y_{_0})\Big].
\end{eqnarray}
By identifying action \eqref{4.6} with the $\sigma$-model
of the form \eqref{a.1} and then by using  \eqref{4.1.3} one can read off the metric and $B$-field corresponding to the action \eqref{4.6}, giving us
\begin{eqnarray}
ds^2 &=& k \Big[d y_{_0}^2 + e^{x_{_1} +2y_{_0}}~ dx_{_1}  dx_{_2} +\frac{1}{4} (dy_{_1}^2+  dx_{_3}^2+ dx_{_4}^2) +\frac{1}{2} \cos y_{_1} dx_{_3} dx_{_4}\Big],\label{4.7}\\
B &=& k  e^{x_{_1} +2y_{_0}}  d{y_{_0}} \wedge  d{x_{_2}} +\frac{k}{4}  \cos {y_{_1}} d{x_{_3}} \wedge d{x_{_4}}.\label{4.7.1}
\end{eqnarray}
Here the coordinates $(y_{_0}, x_{_1}, x_{_2})$ denote the $AdS_3$ space,  while $(x_{_3} , x_{_4}, y_{_1})$
parametrize the sphere $S^3$. One can simply show that under the coordinate transformation
\begin{eqnarray}\label{4.8}
{y_{_0}} = \bar{\phi},~~~~~e^{x_{_1}} = x^+,~~~~{x_{_2}} = x^-,~~~~{x_{_3}} = \psi, ~~~~{x_{_4}} = \varphi, ~~~~{y_{_1}} = \theta,
\end{eqnarray}
the metric \eqref{4.7} and $B$-field \eqref{4.7.1} turn into \eqref{1.1} and \eqref{1.2}, respectively.
Thus, we have constructed a non-Abelian T-dual $\sigma$-model on the $4+2$-dimensional manifold $\M \approx O \times G$ with the Lie group ${A}_2 \otimes 2{A}_1$
whose background of the model describes the $AdS_3 \times S^3$ space. Below,
we construct the non-Abelian T-dual space of this background.


\subsection{The dual $\sigma$-model}

In order to construct the dual $\sigma$-model on the target manifold
$\tilde {{\M}} \approx O \times \tilde {G}$ with the Abelian Lie group  $4A_1$
we parameterize the corresponding Lie
group  with  the coordinates ${\tilde x}_{_{a}} = ({\tilde x}_{_{1}}, \cdots, {\tilde x}_{_{4}})$ so that its element
is defined as \eqref{4.3} by replacing untilded quantities with tilded ones.  Then, by utilizing the commutation relations of \eqref{4.2} and also
formula \eqref{c.8} for tilded quantities we get
\begin{eqnarray}\label{4.9}
{\tilde \Pi}(\tilde g) = \left( \begin{array}{cccc}
                    0 & -{\tilde x}_{_2} & 0 & 0\\
                     {\tilde x}_{_2} & 0 & 0 & 0\\
                      0 & 0 &0 & 0\\
                    0 & 0 & 0 & 0
                      \end{array} \right).
\end{eqnarray}
The dual coupling matrices can be obtained by
inserting \eqref{4.9} and the matrix ${{E_0}_{_{ab}}}$ of \eqref{4.5} into \eqref{c.9}.
They are then read off
{\begin{eqnarray}\nonumber
{\tilde E}^{ab}=\left( \begin{array}{cccc}
                    0 & \frac{1}{{\tilde x}_{_2} + \frac{k}{2} e^{2 y_{_0}}}&0&0\\
                     \frac{1}{-{\tilde x}_{_2} + \frac{k}{2} e^{2 y_{_0}}} & 0&0&0\\
                   0& 0 &\frac{4}{k}&-\frac{8}{k} \cos y_{_1}\\
                  0  & 0 &0& \frac{4}{k}
                      \end{array} \right),~~
                      {\tilde \phi}^{\hspace{0mm}{(1)^{ a}}}_{~~~j} =\left( \begin{array}{cc}
                  -\frac{k e^{2 y_{_0}}}{{\tilde x}_{_2} + \frac{k}{2} e^{2 y_{_0}}} & 0\\
                  0 & 0\\
                  0 & 0\\
                  0 & 0
                      \end{array} \right),
\end{eqnarray}}
\vspace{-2mm}
{\begin{eqnarray}\label{4.10}
{\tilde \phi}_{_{ij}} =\left( \begin{array}{cc}
                    k & 0\\
                  0 & \frac{k}{4}
                      \end{array} \right),~~~~~~~~~~~~~~~~~~~~~~~~~~~~~~
{{\tilde \phi}^{^{{(2)}^{b}}}}_{~i} = \left(\begin{array}{cccc}
                  -\frac{k e^{2 y_{_0}}}{-{\tilde x}_{_2} + \frac{k}{2} e^{2 y_{_0}}} & 0 & 0 & 0 \\
                  0  & 0 & 0 & 0
                      \end{array} \right).
\end{eqnarray}}
Putting these pieces together into \eqref{c.2} and using the fact that the components of the right invariant one-forms on
the dual Lie group, $4A_1$, are ${\tilde R}_{\pm_{a}}=\partial_{\pm} {{\tilde x}_a}$,
the action of dual $\sigma$-model is obtained to be
\begin{eqnarray}\label{4.11}
\tilde S = \frac{1}{2} \int d\sigma^{+}  d\sigma^{-}\hspace{-7mm}&&\Big[k \partial_{+} y_{_0}
\partial_{-} y_{_0} +\frac{k}{4} \partial_{+} y_{_1}  \partial_{-} y_{_1}
+\frac{4}{k} (\partial_{+} {\tilde x_{_3}}  \partial_{-} {\tilde x_{_3}}+\partial_{+} {\tilde x_{_4}}  \partial_{-} {\tilde x_{_4}}) \nonumber\\
&&-\frac{8}{k} \cos y_{_1} ~\partial_{+} {\tilde x_{_3}}  \partial_{-} {\tilde x_{_4}}
+\frac{1}{{\tilde x}_{_2} + \frac{k}{2} e^{2 y_{_0}}} \big(\partial_{+} {\tilde x_{_1}} \partial_{-} {\tilde x_{_2}} -k e^{2 y_{_0}}
\partial_{+} {\tilde x_{_1}} \partial_{-} y_{_0}\big)
\nonumber\\
&&+\frac{1}{-{\tilde x}_{_2} + \frac{k}{2} e^{2 y_{_0}}} \big(\partial_{+} {\tilde x_{_2}} \partial_{-} {\tilde x_{_1}} -k e^{2 y_{_0}}
 \partial_{+} y_{_0} \partial_{-} {\tilde x_{_1}}\big)\Big].
\end{eqnarray}
If one employs the change of coordinates ${\tilde x_{_3}}=-\frac{k}{4} y_{_2}$ and ${\tilde x_{_4}}=\frac{k}{4} y_{_3}$ in the above, then comparing
the resulting action with the $\sigma$-model action of the form \eqref{a.1},
the dual metric and ${\tilde B}$-field take the following forms
\begin{eqnarray}
{\tilde ds}^2 &=& \frac{k  e^{2 y_{_0}}}{\Delta}\big(d {\tilde x_{_1}} d {\tilde x_{_2}}-k  e^{2 y_{_0}} d {\tilde x_{_1}} d {y_{_0}}\big)
 + \frac{k}{4} \big(d {y_{_1}^2} +d {y_{_2}^2}+d {y_{_3}^2}\big) \nonumber\\
&&~~~~~~~~~~~~~~~~~~~~~~~~~~~~~~~~~~~~~~~+k d {y_{_0}^2} +\frac{k}{2} \cos y_{_1}d y_{_2} d y_{_3},\label{4.13}\\
{\tilde  B }&=& \frac{{\tilde x_{_2}}}{\Delta}\big(-d {\tilde x_{_1}}\wedge d {\tilde x_{_2}} +k  e^{2 y_{_0}} d {\tilde x_{_1}} \wedge d {y_{_0}}\big)
+\frac{k}{4}  \cos y_{_1} d y_{_2} \wedge d y_{_3},\label{4.14}
\end{eqnarray}
where $\Delta = {\frac{k^2}{4} e^{4 y_{_0}} -{\tilde x}_{_2}^2}$. Now, one may
introduce the new coordinates $X$, $Y$ and $W$ so that\footnote{Here, the coordinates $y_{_1}$, $y_{_2}$ and $y_{_3}$ remain unchanged.}
\begin{eqnarray}
{{\tilde x}_1} = Y -\frac{k}{2} (e^{^{W}}+W),~~~~{{\tilde x}_2} = k e^X (1+\frac{1}{2} e^{^{-W}}),~~~~
y_{_{0}} = \frac{1}{2} (X-W).\label{4.15}
\end{eqnarray}
If we introduce the new coordinates $({t}, {x}, {r})$  instead of $(X, Y, W)$  by means of the transformation
\begin{eqnarray}
e^{^{W}} =  \frac{1}{r-1},~~~~~X = \frac{2}{\sqrt{k}} (t + \frac{x}{\sqrt{3}}),~~~~~~
Y = \sqrt{k} (t - \frac{x}{\sqrt{3}}),\label{4.16}
\end{eqnarray}
then, \eqref{4.13} and \eqref{4.14} will become, respectively,
\begin{eqnarray}
{d {\tilde s}}^2 &=& - (1-\frac{2}{r}) d t^2 +  (1-\frac{2}{3 r}) d x^2
+\frac{2}{\sqrt{3}}~ d t d x +  \frac{ k}{4 r^2} (1-\frac{1}{r})^{-2} d r^2~~~~~~~~~\nonumber\\
&&~~~~~~~~~~~~~~~~~~~~~~~~~~~~+ \frac{k}{4} \big(d {y_{_1}^2} +d {y_{_2}^2}+d {y_{_3}^2}\big) +\frac{k}{2} \cos y_{_1} d y_{_2} d y_{_3},\label{4.17}\\
{\tilde B} &=& \frac{2}{\sqrt{3}} (1+\frac{1}{r}) dt \wedge dx +\frac{k}{4} \cos y_{_1} d y_{_2} \wedge d y_{_3}.\label{4.18}
\end{eqnarray}
As it can be seen from the metrics \eqref{4.7} and \eqref{4.17}, the duality has changed the $AdS_{3}$ part,
while the $S^{3}$ part has remained unchanged.
Note that one can simply show that for large $r$ the dual part corresponding to the $AdS_{3}$ part of the metric \eqref{4.7} approaches the asymptotically flat solution.
The metric components \eqref{4.17} are ill defined at the regions $r=0$ and $r=1$. We can test whether there are true singularities
by calculating the scalar curvature, which is
\begin{eqnarray}
{\cal R} = \frac{2(3r^2 + 4{r} -7)}{k {{r}^2}}.
\end{eqnarray}
Thus, $ r=0 $ is a true singularity; moreover, one can show that this
singularity also appears in the Kretschmann scalar, which is, ${\mathcal{K}}=4(3r^4 +8r^2-24 {r} +19)/k^2 {r}^4$.
Also, $r=1$ can be a single horizon. According to the above result, the non-Abelian T-duality transformation has been related a solution with no horizon and no
curvature singularity to a solution with a single horizon and a curvature singularity.


\section{Integrability of the T-dual $\sigma$-models}
\label{section4}
Before proceeding to investigate the integrability of the T-dual $\sigma$-models built in the previous section,
let us review the Lax formulation of integrability
of a prescription invented by N. Mohammedi \cite{N.Mohammedi}.
The Lax formulation of integrability provides a method for constructing conserved dynamical quantities.
According to \cite{N.Mohammedi},  a 2-dimensional  $\sigma$-model is classically integrable if its equations of motion can be represented as
a zero curvature relation. This means that a Lax pair $({\cal A}_+(\lambda)~,~{\cal A}_-(\lambda))$
can be found for all values of the spectral parameter $\lambda$ such that the commutator
\begin{eqnarray}
[\partial_+ + {\cal A}_+(\lambda) ~,~\partial_- +{\cal A}_-(\lambda)]=0,\label{44.1}
\end{eqnarray}
yields the equations of motion of the $\sigma$-model under consideration.

In the absence of a dilaton field, the action of $\sigma$-model \eqref{a.1} in the standard light-cone coordinates may be expressed as
\begin{eqnarray}
S =\frac{1}{2}\int_{{\Sigma}} d\sigma^+d\sigma^-  ({G}_{_{MN}} + {B}_{_{MN}})  \partial_+ X^{^M} \partial_-X^{^N}, \label{44.2}
\end{eqnarray}
The equations of motion of this action can be written as
\begin{eqnarray}
 \partial_+ \partial_-X^{^M} + (\Gamma^M_{~_{NP}}- H^M_{~_{NP}})\partial_+X^{^N} \partial_-X^{^P}=0, \label{44.3}
\end{eqnarray}
where $\Gamma^M_{~_{NP}}$'s are conventional Christoffel symbols, and $H^M_{~_{NP}} = G^{^{MQ}} H_{_{QNP}}$ such that $H_{_{MNP}}$ defined by
\begin{eqnarray}
H_{_{MNP}}=\frac{1}{2}(\partial_{_M} B_{_{NP}}+\partial_{_N} B_{_{PM}}+\partial_{_P} B_{_{MN}}), \label{44.4}
\end{eqnarray}
is the field strength of $B$-field.
Let us now construct a linear system whose consistency conditions are equivalent to the
equations of motion \eqref{44.3}. To this end, we define the Lax pair  $({\cal A}_+(\lambda)~,~{\cal A}_-(\lambda))$ as
$( \partial_+ X^{^M} \alpha_{_M}(\lambda , X) ,  \partial_- X^{^N}\beta_{_N} (\lambda , X))$, and as an ansatz we take the following
linear system
\begin{eqnarray}
{[\partial_+ + \partial_+ X^{^M} \alpha_{_M}(\lambda, X)]} \Psi &=&0,\nonumber\\
{[\partial_- + \partial_- X^{^N} \beta_{_N} (\lambda , X)]} \Psi &=& 0,\label{44.5}
\end{eqnarray}
where the matrices $\alpha_{_M}(\lambda, X)$ and $\beta_{_N} (\lambda , X)$ depend on the fields $ X^{^M}$ and
possibly on some free arbitrary parameter $\lambda$.
The arbitrary field $\Psi$ can be a column vector.
The compatibility condition of the linear system \eqref{44.5} yields the equations of motion if the matrices
$\alpha_{_M}(\lambda, X)$ and $\beta_{_N} (\lambda , X)$ satisfy the following relation \cite{N.Mohammedi}
\begin{eqnarray}
\partial_{_M} \beta_{_N} - \partial_{_N} \alpha_{_M}  + [\alpha_{_M} , \beta_{_N}] =  (\Gamma^P_{~_{MN}}- H^P_{~_{MN}})\lambda_{_P},\label{44.6}
\end{eqnarray}
where we have defined  $\lambda_{_M} := \beta_{_M}-\alpha_{_M}$. By using this, one may rewrite the equation \eqref{44.6} in the following form
\begin{eqnarray}
\partial_{_M} \alpha_{_N} - \partial_{_N} \alpha_{_M}  + [\alpha_{_M} , \alpha_{_N}] + \partial_{_M} \lambda_{_N}
+ [\alpha_{_M} , \lambda_{_N}]=  (\Gamma^P_{~_{MN}}- H^P_{~_{MN}})\lambda_{_P}.\label{44.7}
\end{eqnarray}
By splitting the above equation  into its symmetric and anti-symmetric parts, one can express it
as the following set of relations
\begin{eqnarray}
{\partial_{_M} \lambda_{_N} + \partial_{_N} \lambda_{_M}  + [\alpha_{_M} , \lambda_{_N}]
+ [\alpha_{_N} , \lambda_{_M}]  - 2\Gamma^P_{~_{MN}} \lambda_{_P}} &=& 0,~~~~~\label{44.8}\\
\partial_{_M} \alpha_{_N} - \partial_{_N} \alpha_{_M}  + [\alpha_{_M} , \alpha_{_N}] +\frac{1}{2}(\partial_{_M} \lambda_{_N}
-\partial_{_N} \lambda_{_M})~~~~~~~~~~~~~~~~~~~~~~~&&\nonumber\\
~~~~~~~~~~~~~~~~~~~~~~~~~~~~~~~~~~~~~~~~~~+
 \frac{1}{2}\big([\alpha_{_M} , \lambda_{_N}]- [\alpha_{_N} , \lambda_{_M}]\big)+ H^P_{~_{MN}}\lambda_{_P} &=& 0.\label{44.9}
\end{eqnarray}
Thus, the integrability condition of the $\sigma$-model \eqref{44.2} is equivalent to finding matrices $\alpha_{_M}$ and $\lambda_{_N}$
that satisfy the pair of equations \eqref{44.8} and \eqref{44.9}.
Note that these equations are at the center of the ability to represent the equations of motion of
the $\sigma$-model \eqref{44.2} as a zero curvature condition of a linear system. The unknowns
of the problem are the two sets of matrices $\alpha_{_M}$ and $\lambda_{_M}$ and also the Christoffel symbols $\Gamma^P_{~_{MN}}$
and field strength $H^P_{~_{MN}}$.
Before proceeding to find the matrices $\alpha_{_M}$ and $\lambda_{_M}$ from equations \eqref{44.8} and \eqref{44.9}, one must first
extract $\Gamma^P_{~_{MN}}$ and $H^P_{~_{MN}}$ from the knowledge of $G_{_{MN}}$ and  $B_{_{MN}}$ of a given $\sigma$-model.
\\\\
$\bullet$~{\bf Investigating the integrability of the original $\sigma$-model.}
The line element and $B$-field of the original $\sigma$-model \eqref{4.6} have been presented in equations \eqref{4.7}
and \eqref{4.7.1}, respectively. In this manner, one can extract the corresponding  Christoffel symbols and field strength.
Thus, the only remaining unknowns of the problem will be matrices $\alpha_{_M}$ and $\lambda_{_M}$.
In order to find some solutions, we proceed by fixing some of these unknowns.
We take the following expressions for the matrices $\alpha_{_M}$ and $\lambda_{_M}$
\begin{eqnarray}
\alpha_{_M} = {R_{_M}}^a  ~ {A_{_a}}^{^b}~ T_{_b},~~~~~~~~\lambda_{_M} ={R_{_M}}^a ~ {C_{_a}}^{^b}~ T_{_b}, \label{44.10}
\end{eqnarray}
where ${R_{_M}}^a$ are the components of the right-invariant Maurer-Cartan one-forms which have defined in section \ref{section2}, and
 ${T}_a$, $a=1, \cdots, dim~G$ are the bases of the Lie algebra ${\G}$ of ${G}$.
Here, the quantities ${A_{_a}}^{^b}$ and ${C_{_a}}^{^b}$ are two constant square matrices which
will provide the spectral parameter.
Injecting the expressions of $\alpha_{_M}$ and $\lambda_{_M}$ in equations \eqref{44.8} and \eqref{44.9} leads to
\begin{eqnarray}
\big(\partial_{_M} {R_{_N}}^a + \partial_{_N} {R_{_M}}^a\big) {C_{_a}}^{^e}+  \big({R_{_M}}^a {R_{_N}}^d + {R_{_N}}^a {R_{_M}}^d\big)
 {A_{_a}}^{^b}  {C_{_d}}^{^c}  f^e_{~bc} -2 \Gamma^{^{P}}_{~_{MN}} {R_{_P}}^b {C_{_b}}^{^e} &=& 0,~~~~~~~\label{44.11}\\
- \big({A_{_a}}^{^e} +\frac{1}{2} {C_{_a}}^{^e}\big)  {R_{_M}}^b {R_{_N}}^c  f^a_{~bc}  +\frac{1}{2}\big({R_{_M}}^a {R_{_N}}^d - {R_{_N}}^a {R_{_M}}^d\big)
{A_{_a}}^{^b} {C_{_d}}^{^c} f^e_{~bc} \nonumber\\
+ {R_{_M}}^a {R_{_N}}^d {A_{_a}}^{^b} {A_{_d}}^{^c} f^e_{~bc}+H^{^{P}}_{~_{MN}} {L_{_P}}^b {C_{_b}}^{^e} &=& 0.\label{44.12}
\end{eqnarray}
In obtaining equation \eqref{44.12} we have used
the fact that the ${R_{_M}}^a$'s satisfy the Maurer-Cartan equation
$\partial_{_M} {R_{_N}}^a - \partial_{_N} {R_{_M}}^a=-f^a_{~bc} {R_{_M}}^b {R_{_N}}^c $.

In order to obtain the matrices $\alpha_{_M}$ and $\lambda_{_M}$ from equations \eqref{44.11} and \eqref{44.12}, one first
finds that the only non-zero components of the Christoffel symbols corresponding to metric \eqref{4.7} are
\begin{eqnarray}
{\Gamma^{^{x_{_1}}}_{~_{x_{_1} x_{_1}}}} &=& \Gamma^{^{x_{_1}}}_{~_{x_{_1} y_{_0}}}=\Gamma^{^{x_{_2}}}_{~_{x_{_2} y_{_0}}} =1,
~~~~~\Gamma^{^{x_{_3}}}_{~_{x_{_4} y_{_1}}}
=\Gamma^{^{x_{_4}}}_{~_{x_{_3} y_{_1}}} = -\frac{1}{2} \csc y_{_1},~~~~\nonumber\\
{\Gamma^{^{x_{_3}}}_{~_{x_{_3} y_{_1}}}} &=&  \Gamma^{^{x_{_4}}}_{~_{x_{_4} y_{_1}}} = \frac{1}{2} \cot y_{_1},~~~~~~\Gamma^{^{y_{_0}}}_{~_{x_{_1} x_{_2}}} = -\frac{1}{2} e^{x_{_1} + 2 y_{_0}},~~~~\Gamma^{^{y_{_1}}}_{~_{x_{_3} x_{_4}}} =
\frac{1}{2} \sin y_{_1}.~~ \label{44.13}
\end{eqnarray}
It is also straightforward to verify that the only non-zero components of the strength field corresponding to $B$-field \eqref{4.7.1} are
\begin{eqnarray}
{H^{^{x_{_1}}}_{~_{x_{_1} y_{_0}}}} &=& H^{^{x_{_2}}}_{~_{ y_{_0} x_{_2}}} =1,
~~~~~~~~~~~~~~H^{^{x_{_3}}}_{~_{y_{_1} x_{_3} }}
=H^{^{x_{_4}}}_{~_{x_{_4} y_{_1}}} = \frac{1}{2} \cot y_{_1},~~~~\nonumber\\
{H^{^{x_{_3}}}_{~_{ y_{_1} x_{_4}}}} &=& H^{^{x_{_4}}}_{~_{x_{_3} y_{_1}}} = \frac{1}{2}\csc y_{_1},~~~H^{^{y_{_0}}}_{~_{x_{_1} x_{_2}}} = -\frac{1}{2} e^{x_{_1} + 2 y_{_0}},~~~H^{^{y_{_1}}}_{~_{x_{_3} x_{_4}}} =
-\frac{1}{2} \sin y_{_1}.~~ \label{44.14}
\end{eqnarray}
Here, the Lie algebra $\G$ is considered to be ${\A}_2 \oplus 4{\A}_1$ which is defined by the following
commutation relations\footnote{In the previous section, we showed that the $\sigma$-model \eqref{4.6}
was built on a $4+2$-dimensional manifold $\M \approx {\cal O} \times G$
 with the 4-dimensional Lie group ${A}_2 \otimes 2{A}_1$ and spectator fields $(y_{_0} , y_{_1})$.
We furthermore mentioned that $R_{\pm}^i =\partial_{\pm} y^{i}$, and this shows that spectator fields can play the role of Abelian generators of Lie algebra.
Accordingly, in order to check the integrability of the $\sigma$-model \eqref{4.6}, we have considered the Lie group ${A}_2 \otimes 4{A}_1$ instead of the $4+2$-dimensional manifold $\M$.}
\begin{eqnarray}\label{44.15}
[T_{_1} , T_{_2}] = T_2,~~~~[T_{_3}~ , ~.]~=~0,~~~~[T_{_4} ~, ~.]=0,~~~~[T_{_5}~ , ~.]~=~0,~~~~[T_{_6} ~, ~.]=0.
\end{eqnarray}
In order to calculate the ${R_{_M}}^a$'s on the ${A}_2 \otimes 4{A}_1$, we choose
a convenient element of the group manifold. Then we obtain $R_{\pm}^2 =e^{x_{_1}}\partial_{\pm} x_{_2}, R_{\pm}^a =\partial_{\pm} x_{_a},~a=1, 3, 4, 5, 6$.
By putting these into equations \eqref{44.11} and \eqref{44.12} one obtains the constant matrices ${A_{_a}}^{^b}$ and ${C_{_a}}^{^b}$, giving us
\begin{eqnarray}\label{44.16}
{A_{_a}}^{^b}=\left( \begin{array}{cccccc}
                    -$1$ & $$\gamma_{_1}$$ & $$\gamma_{_2}$$ & $$\gamma_{_3}$$ & $$\gamma_{_4}$$ & $$\gamma_{_5}$$\\
                     0 & 0 & 0 & 0 & 0 & 0\\
                   0 & 0 & $$\xi_{_1}$$ & $$\xi_{_2}$$ & $$\xi_{_3}$$ & $$\xi_{_4}$$\\
                  0 & 0 & $$\zeta_{_1}$$ & $$\zeta_{_2}$$ & $$\zeta_{_3}$$ & $$\zeta_{_4}$$\\
                   -2 & $$\eta_{_1}$$ & $$\eta_{_2}$$ & $$\eta_{_3}$$ & $$\eta_{_4}$$ & $$\eta_{_5}$$\\
                    0 & 0 & $$\rho_{_1}$$ & $$\rho_{_2}$$ & $$\rho_{_3}$$ & $$\rho_{_4}$$
                      \end{array} \right),~~~{C_{_a}}^{^b}=\left( \begin{array}{cccccc}
                    0 & $$2 \gamma_{_1}-\eta_{_1}$$ & 0 & 0 & 0 & 0\\
                     0 & 0 & 0 & 0 & 0 & 0\\
                   0 & 0 & 0 & 0 & 0 & 0\\
                  0 & 0 & 0 & 0 & 0 & 0\\
                   0 & 0 & 0 & 0 & 0 & 0\\
                    0 & 0 & 0 & 0 & 0 & 0
                      \end{array} \right),
\end{eqnarray}
where $\gamma_{_i}$, $\xi_{_i}$, $\zeta_{_i}$, $\eta_{_i}$ and $\rho_{_i}$ are some arbitrary parameters.
Thus, the Lax pair of the integrable non-linear $\sigma$-model \eqref{4.6} is of the form \eqref{44.5} with
$\alpha_{_M} = {R_{_M}}^a  ~ {A_{_a}}^{^b}~ T_{_b}$ and $\beta_{_M} = {R_{_M}}^a  ({A_{_a}}^{^b}+ {C_{_a}}^{^b})~ T_{_b}$.
\\\\
$\bullet$~{\bf Investigating the integrability of the dual $\sigma$-model.}
We will show now that if the original theory \eqref{4.6} is integrable then its dual (background given by relations \eqref{4.17} and \eqref{4.18}) is also
integrable.
Since we were dealing with the non-Abelian T-duality, the Lie group of the dual target manifold is considered to be Abelian Lie group $6 A_1$.
Accordingly, the right-invariant one-forms on the $6 A_1$ are ${\tilde R}_{\pm_{a}} =
\partial_{\pm} {\tilde X}^{^M} {\tilde R}_{_{Ma}}= \partial_{\pm} {\tilde x}_{a}$, in which ${\tilde x}_{a}, a=1, \cdots, 6,$ stand for the coordinates of the
$6 A_1$.
In order to investigate the integrability of the dual $\sigma$-model, the corresponding linear system is taken to have the form \eqref{44.5} by
replacing the untilded symbols by tilded ones.
For matrices ${\tilde \alpha}_{_M}$ and ${\tilde \lambda}_{_M}$, one may consider the following expansions
\begin{eqnarray}
{\tilde \alpha}_{_M} = {\tilde R}_{_{Ma}} ~ {\tilde A^{^a}}_{_{~b}}({\tilde x})~ {\tilde T}^{^b},~~~~~~~~
{\tilde \lambda}_{_M} ={\tilde R}_{_{Ma}} ~ {\tilde C^{^a}}_{_{~b}}({\tilde x})~ {\tilde T}^{^b}.\label{44.17}
\end{eqnarray}
Here we have assumed that the ${\tilde A^{^a}}_{_{~b}}({\tilde x})$ and ${\tilde C^{^a}}_{_{~b}}({\tilde x})$ are not constant and depend on the coordinates of the group manifold. Inserting the above expansion into equations \eqref{44.8} and \eqref{44.9} and using the fact that
${{\tilde f}^{ab}}_{\; \; \: c} =0$, one gets
\begin{eqnarray}
{\tilde R}_{_{Ma}}  {\tilde \partial}_{_N} {\tilde C^{^a}}_{_{~b}}({\tilde x}) + {\tilde R}_{_{Na}}  \partial_{_M} {\tilde C^{^a}}_{_{~b}}({\tilde x})
-2 {\tilde \Gamma}^{^{P}}_{~_{MN}} {\tilde R}_{_{Pa}} {\tilde C^{^a}}_{_{~b}}({\tilde x}) &=& 0,~~~~~~~\label{44.18}\\
\big[\partial_{_M} {\tilde A^{^a}}_{_{~b}}({\tilde x}) +\frac{1}{2} \partial_{_M} {\tilde C^{^a}}_{_{~b}}({\tilde x})\big]
{\tilde R}_{_{Na}} -{\tilde R}_{_{Ma}} \big[\partial_{_N} {\tilde A^{^a}}_{_{~b}}({\tilde x}) +\frac{1}{2} \partial_{_N} {\tilde C^{^a}}_{_{~b}}({\tilde x})\big]
\nonumber\\
+{\tilde H}^{^{P}}_{~_{MN}} {\tilde R}_{_{Pa}} {\tilde C^{^a}}_{_{~b}}({\tilde x}) &=& 0.\label{44.19}
\end{eqnarray}
The Christoffel symbols ${\tilde \Gamma}^{^{P}}_{~_{MN}}$ and strength field ${\tilde H}^{^{P}}_{~_{MN}}$ are those corresponding to the
metric ${\tilde G}_{_{MN}}$ and anti-symmetric tensor ${\tilde B}_{_{MN}}$ of equations \eqref{4.17} and \eqref{4.18}, respectively.
Hence, one obtains that
\begin{eqnarray}
{{\tilde \Gamma}^{^r}_{~_{rr}}} &=& \frac{1}{1-r},~~~~~~~~~~~~~~~~~{\tilde \Gamma}^{^{r}}_{~_{tt}}= -3{\tilde \Gamma}^{^{r}}_{~_{xx}} =\frac{4(r-1)^2}{k r^2},
~~~~~~~~~~{\tilde \Gamma}^{^{t}}_{~_{rt}}=\frac{-2+3r}{4 r (r-1)^2},~~\nonumber\\
{{\tilde \Gamma}^{^{t}}_{~_{rx}}} &=& -\frac{1}{3} {{\tilde \Gamma}^{^{x}}_{~_{rt}}} = \frac{1}{4 \sqrt{3} (r-1)^2},~~~~  {{\tilde \Gamma}^{^{x}}_{~_{rx}}} = \frac{r-2}{4 r  (r-1)^2},~~~~~~~~~{{\tilde \Gamma}^{^{y_{_1}}}_{~_{y_{_2} y_{_3}}}} = \frac{1}{2} \sin y_{_1}, \nonumber\\
{{\tilde \Gamma}^{^{y_{_2}}}_{~_{y_{_1} y_{_2}}}}&=& {{\tilde \Gamma}^{^{y_{_3}}}_{~_{y_{_1} y_{_3}}}} =\frac{1}{2}
\cot y_{_1},~~~~~~~~{{\tilde \Gamma}^{^{y_{_2}}}_{~_{y_{_1} y_{_3}}}} = {{\tilde \Gamma}^{^{y_{_3}}}_{~_{y_{_1} y_{_2}}}} =
- \frac{1}{2} \csc y_{_1},\label{44.20}
\end{eqnarray}
and
\begin{eqnarray}
{\tilde H^{^{r}}_{~_{t x}}} &=& -\frac{4(r-1)^2}{\sqrt{3} k r^2},~~~~~~~~~~~~~~~~~{\tilde H^{^{x}}_{~_{rt}}}= -\frac{\sqrt{3}}{4  (r-1)^2} (1-\frac{2}{r}),
~~~~~~{\tilde H^{^{t}}_{~_{rx}}}= \frac{2-3 r}{4\sqrt{3} r  (r-1)^2},\nonumber\\
{\tilde H^{^{x}}_{~_{rx}}} &=& - {\tilde H^{^{t}}_{~_{rt}}} =\frac{1}{4(r-1)^2},~~~~~{\tilde H^{^{y_{_2}}}_{~_{y_{_1} y_{_3}}}} = -{\tilde H^{^{y_{_3}}}_{~_{y_{_1} y_{_2}}}} = \frac{1}{2} \csc y_{_1},\nonumber\\
{\tilde H^{^{y_{_2}}}_{~_{y_{_1} y_{_2}}}} &=& -{\tilde H^{^{y_{_3}}}_{~_{y_{_1} y_{_3}}}} = \frac{1}{2} \cot y_{_1},~~~{\tilde H^{^{y_{_1}}}_{~_{y_{_2} y_{_3}}}} = -\frac{1}{2} \sin y_{_1}.
\label{44.21}
\end{eqnarray}
Finally, inserting \eqref{44.20} and \eqref{44.21} into equations \eqref{44.18} and \eqref{44.19} and using
the fact that ${\tilde R}_{_{Ma}} = {\delta}_{_{Ma}}$, one can find the matrices ${\tilde A^{^a}}_{_{~b}}({\tilde x})$ and
${\tilde C^{^a}}_{_{~b}}({\tilde x})$. The result is
\begin{eqnarray}\nonumber
{\tilde A^{^a}}_{_{~b}}({\tilde x})=\left( \begin{array}{cccccc}
                   $${\tilde \xi}_{_1}$$ & $${\tilde \xi}_{_2}$$ & $${\tilde \xi}_{_3}$$ & $${\tilde \xi}_{_4}$$ & $${\tilde \xi}_{_5}$$ & $${\tilde \xi}_{_6}$$\\
                    $$-\frac{\sqrt{3} }{r} {\tilde \eta}_{_1}$$ & $$-\frac{\sqrt{3} }{r} {\tilde \eta}_{_2}$$ &$$-\frac{\sqrt{3} }{r} {\tilde \eta}_{_3}$$ & $$-\frac{\sqrt{3} }{r} {\tilde \eta}_{_4}$$ & $$-\frac{\sqrt{3} }{r} {\tilde \eta}_{_5}$$ & $$-\frac{\sqrt{3} }{r} {\tilde \eta}_{_6}$$\\
                   $$\frac{2}{r} {\tilde \eta}_{_1}$$ & $$\frac{2}{r} {\tilde \eta}_{_2}$$ &$$\frac{2}{r}{\tilde \eta}_{_3}$$ & $$\frac{2}{r}{\tilde \eta}_{_4}$$ & $$\frac{2}{r}{\tilde \eta}_{_5}$$ & $$\frac{2}{r}{\tilde \eta}_{_6}$$\\
                   $${\tilde \zeta}_{_1}$$ & $${\tilde \zeta}_{_2}$$ & $${\tilde \zeta}_{_3}$$ & $${\tilde \zeta}_{_4}$$ & $${\tilde \zeta}_{_5}$$ & $${\tilde \zeta}_{_6}$$\\
                    $${\tilde \gamma}_{_1}$$ & $${\tilde \gamma}_{_2}$$ & $${\tilde \gamma}_{_3}$$ & $${\tilde \gamma}_{_4}$$ & $${\tilde \gamma}_{_5}$$ & $${\tilde \gamma}_{_6}$$\\
                    $$\frac{1}{2}{\tilde \mu}_{_1} \cos y_{_1}$$& $$\frac{1}{2}{\tilde \mu}_{_2} \cos y_{_1}$$ & $$\frac{1}{2}{\tilde \mu}_{_3} \cos y_{_1}$$ & $$\frac{1}{2}{\tilde \mu}_{_4} \cos y_{_1}$$ & $$\frac{1}{2}{\tilde \mu}_{_5} \cos y_{_1}$$ & $$\frac{1}{2}{\tilde \mu}_{_6} \cos y_{_1}$$
                      \end{array} \right),
\end{eqnarray}
\vspace{-3mm}
\begin{eqnarray}\label{44.16}
                      {\tilde C^{^a}}_{_{~b}}({\tilde x})=\left( \begin{array}{cccccc}
                    0 & 0 & 0 & 0 & 0 & 0\\
                    $$f(r) {\tilde \eta}_{_1}$$ & $$f(r) {\tilde \eta}_{_2}$$ & $$f(r) {\tilde \eta}_{_3}$$ & $$f(r) {\tilde \eta}_{_4}$$ & $$f(r) {\tilde \eta}_{_5}$$ & $$f(r) {\tilde \eta}_{_6}$$\\
                    $${\tilde \eta}_{_1}$$ & $${\tilde \eta}_{_2}$$ & $${\tilde \eta}_{_3}$$ & $${\tilde \eta}_{_4}$$ & $${\tilde \eta}_{_5}$$ & $${\tilde \eta}_{_6}$$\\
                    0 & 0 & 0 & 0 & 0 & 0\\
                    $${\tilde \mu}_{_1}$$ & $${\tilde \mu}_{_2}$$ & $${\tilde \mu}_{_3}$$ & $${\tilde \mu}_{_4}$$ & $${\tilde \mu}_{_5}$$ & $${\tilde \mu}_{_6}$$\\
                    $${\tilde \mu}_{_1} \cos y_{_1}$$& $${\tilde \mu}_{_2} \cos y_{_1}$$ & $${\tilde \mu}_{_3} \cos y_{_1}$$ & $${\tilde \mu}_{_4} \cos y_{_1}$$ & $${\tilde \mu}_{_5} \cos y_{_1}$$ & $${\tilde \mu}_{_6} \cos y_{_1}$$
                      \end{array} \right), ~~
\end{eqnarray}
where $f(r)=-\sqrt{3}  (1-\frac{2}{r})$, and the parameters ${\tilde \eta}_{_i}$, ${\tilde \mu}_{_i}$, ${\tilde \gamma}_{_i}$, ${\tilde \xi}_{_i}$ and ${\tilde \zeta}_{_i}$ are arbitrary.

\section{Conformality of the T-dual $\sigma$-models}
\label{section5}
Before proceeding to investigate the conformal invariance conditions of the $AdS_3 \times S^3$ background and its dual pair, let us
introduce the vanishing of the beta-function equations up to two-loop order.
Consistency of the string theory requires that the action \eqref{a.1}
defines a conformally invariant quantum field theory. The conformal invariance conditions can be interpreted as effective field equations for the coupling functions $G_{_{MN}}(X)$, $B_{_{MN}}(X)$ and $\Phi(X)$
of the string effective action \cite{callan}.
We note that the conditions for conformal
invariance at the quantum level, which is equivalent to the vanishing of the beta functions, and the requirement
of integrability of T-dual $\sigma$-models might reduce the number of possibilities
for the spaces on which one can carry out the compactification of the extra dimensions of string theory.

In the $\sigma$-model context, the conformal invariance conditions of the $\sigma$-model \eqref{a.1} are
provided by the vanishing of the beta-function equations \cite{callan}.
At the two-loop level (first order in $\alpha'$) these equations read \cite{{A.Sen1},{Tseytlin},{c.hull},{Metsaev}}
\begin{eqnarray}
{\beta_{_{MN}}^G}&=& {{\beta_{_{MN}}^G}}^{\hspace{-2mm}(1)} +\alpha' ~ {{\beta_{_{MN}}^G}}^{\hspace{-2mm}(2)}+{\cal O}(\alpha'^2) \nonumber\\
	&=&{\cal R}_{_{MN}}-H^2_{_{MN}}+{\nabla}_{_M}
{\nabla}_{_N} \Phi +\frac{1}{2} \alpha' \Big[{\cal R}_{_{M PQR}} {\cal R}_{_N}^{^{~PQR}}
+2 {\cal R}_{_{M PQ N}} {H^2}^{^{PQ}}\nonumber\\
&&~~~~~+2 {\cal R}_{_{PQ R(M}}H_{_N)}^{^{~R S}} H^{^{PQ}}_{~~_{S}} +\frac{1}{3} ({\nabla}_{_M} H_{_{PQR}})
({\nabla}_{_N} H^{^{PQR}})-({\nabla}_{_P} H_{_{RS M}})
({\nabla}^{^P} H^{^{RS}}_{~~_{N}})\nonumber\\
&&~~~~~+2 H_{_{MPQ}} H_{_{N R S}} H^{^{T S Q}} H_{_T}^{^{~~R P}}
+2 H_{_{M P Q}} H_{_{NR}} ^{^{~~Q}}   {H^2}^{^{R P}} \Big]+
{\cal O}(\alpha'^2)~=~0,\label{5.1}
\end{eqnarray}
\vspace{-6mm}
\begin{eqnarray}
{\beta_{_{MN}}^B}&=& {{\beta_{_{MN}}^B}}^{\hspace{-2mm}(1)} +\alpha' ~ {{\beta_{_{MN}}^B}}^{\hspace{-2mm}(2)}+{\cal O}(\alpha'^2)~~~~~~~~~~~~~ \nonumber\\
	&=&{\nabla}^{^P} H_{_{PMN}} -  ({\nabla}^{^P}\Phi')  H_{_{MNP}}
+\alpha' \Big[{\nabla}^{^P} H^{^{RS}}_{_{~~[M}}{\cal R}_{_{_{{N]} P RS}}} -({\nabla}_{_P} H_{_{R MN}}) {H^2}^{^{PR}}\nonumber\\
&&~~~-2 ({\nabla}^{^P}H^{^{Q R}}_{_{~~[M}})H_{_{_{{N]} Q S }}} H_{_{P R}}^{^{~\;S}}\Big]
+{\cal O}(\alpha'^2)~=~0,~~~~~~~\label{5.2}
\end{eqnarray}
\vspace{-6mm}
\begin{eqnarray}
{\beta^\Phi}&=& {{\beta^\Phi}}^{(1)} +\alpha' ~ {{\beta^\Phi}}^{(2)}+{\cal O}(\alpha'^2)\nonumber\\
	&=&2 \Lambda + {\nabla}^2 \Phi' - ({\nabla} \Phi')^2+\frac{2}{3} H^{{2}}
-\alpha' \Big[\frac{1}{4} {\cal R}_{_{MNRS}} {\cal R}^{^{MNRS}}
-\frac{1}{3} ({\nabla}_{_M} H_{_{NRS}})
 ({\nabla}^{^M} H^{^{NRS}})
\nonumber\\
&&~~~~~-\frac{1}{2} H^{^{MN}}_{_{~~P}} H^{^{RS P}} {\cal R}_{_{MN RS}}-{\cal R}_{_{MN}} {H^2}^{^{MN}} +\frac{3}{2} H^2_{_{MN}} {H^2}^{^{MN}}\nonumber\\
&&~~~~~+\frac{5}{6} H_{_{MN P }} H^{^M}_{_{~~RS}} H^{^{N R}}_{_{~~Q}} H^{^{P S Q}}\Big] +{\cal O}(\alpha'^2)=0,\label{5.3}
\end{eqnarray}
where the field strength $H_{_{MNP}}$ has already defined in equation \eqref{44.4}.
We have introduced the conventional notations $H^2_{_{MN}}=H_{_{MPQ}}  H^{^{PQ}}_{_{~~N}}$, $H^2=H_{_{MNP}}
H^{^{MNP}}$, ${H^2}^{^{MN}} = H^{^{MPQ}} H_{_{PQ}}^{^{~~N}}$ and $({\nabla} \Phi)^2 =\partial_{_{M}} \Phi ~\partial^{^{M}} \Phi$. Furthermore,
${\cal R}_{_{MN}}$ and ${\cal R}_{_{MNPQ}}$ are the Ricci tensor and
Riemann tensor field of the metric
$G_{_{MN}}$, respectively.
We note that in equation \eqref{5.3}, $\Phi' = \Phi + \alpha' q H^2$ for some coefficient $q$ \cite{c.hull}, and
$\Lambda$ is the cosmological constant; moreover, the round brackets denote the symmetric part on the indicated indices whereas square brackets
denote the anti-symmetric part.
Below we shall show that the $AdS_3 \times S^3$ background can be considered as a solution in string theory for the full ${\cal O}(\alpha')$ action
including both dilaton and axion fields. Moreover, it is shown that the dual background (the metric \eqref{4.17} and $B$-field \eqref{4.18})
remains conformal up to one-loop order only.

\subsection{Solutions up to one-loop order, zeroth order in $\alpha'$}

As mentioned in section \ref{section2}, the dilaton field that makes the original $\sigma$-model conformal
up to the one-loop order must obey equation \eqref{c.16.1}. On the other hand, since $\Pi=0$, it follows from the first equation of \eqref{c.7}
that $E=E_0$, and thus from  \eqref{c.16.1} we have $\Phi =\varphi^{^{(0)}} - \log |\det {a(g)}|$.
To obtain the matrix $a(g)$ we use equations \eqref{c.8} and \eqref{4.1}. Then, one gets $\det a(g) =e^{-x_1}$, and hence
$\Phi =\varphi^{^{(0)}} + x_1$. In order to satisfy equations \eqref{5.1}-\eqref{5.3} up to one-loop order (${{\beta_{_{MN}}^G}}^{\hspace{-2mm}(1)}=0,
{{\beta_{_{MN}}^B}}^{\hspace{-2mm}(1)}=0, {{\beta^\Phi}}^{(1)}=0$)
with the metric \eqref{4.7}
and $B$-field \eqref{4.7.1},
since we want the total dilaton to be constant, ${\Phi} = c_{_0}$, we need to
choose $\varphi^{^{(0)}} =c_{_0} -x_1$. Finally, we conclude that the $AdS_3 \times S^3$ background including
the metric \eqref{4.7} and $B$-field \eqref{4.7.1} is conformally invariant up to one-loop order with a constant dilaton field,
in such a way that the cosmological constant is obtained to be zero.

Let us turn our attention to the dual model. The new dilaton field that makes the dual $\sigma$-model \eqref{4.11} conformal is obtained
by solving equations \eqref{5.1}-\eqref{5.3} up to one-loop order, giving\footnote{Note that the dilatonic contribution in \eqref{5.3} up to one-loop order
is vanished if the cosmological constant be zero.}
\begin{eqnarray}\label{5.5}
{\tilde \Phi} = c_{_0}-\log|\frac{2{\tilde x}_{_2} + {k} e^{2 y_{_0}}}{2{\tilde x}_{_2} - k e^{2 y_{_0}}}|.
\end{eqnarray}
In the case of the transformed background  (the metric \eqref{4.17} and $\tilde B$-field \eqref{4.18}), this field is ${\tilde \Phi} = c_{_0}-\log r$.
It should be noted that the dilaton field obtained in \eqref{5.5} does not follow the transformation \eqref{c.16.2}.
The reason behind this may be due to the fact that models defined for Manin triples whose adjoint representations corresponding to structure
coefficients (${{({\cal X}_b)}_c}^{a}=-{f^a}_{bc}$) have non-zero trace  (which is the case of the
semi-Abelian double $({\A}_2 \oplus 2{\A}_1 , 4{\A}_1)$), are anomalous
on the quantum level. For the ${\A}_2 \oplus 2{\A}_1$ Lie algebra, one gets tr${({\cal X}_1)}=-1$, tr${({\cal X}_i)}= 0,~ i=2,3,4$.
In this case, the effective action has a gravitational anomaly that
cannot be absorbed into the transformation of the dilaton.
As shown in \cite{N.Mohammedi2}, a sufficient condition for the invariance of
the reduced string effective action under PL T-duality is the vanishing of the trace of the
adjoint representations of each Lie algebra making up the Drinfeld double.

\subsection{Solutions up to two-loop order, first order in $\alpha'$}

For the $AdS_3 \times S^3$ background including
the metric \eqref{4.7} and $B$-field \eqref{4.7.1} with a constant dilaton field, the field equations \eqref{5.1} and \eqref{5.2} up to two-loop order are fulfilled. Also, the dilatonic contribution in \eqref{5.3}
is vanished   if the following condition holds between the constants $k, \alpha'$  and $\Lambda$:
\begin{eqnarray}\label{5.6}
\Lambda -\frac{8 \alpha'}{k^2} =0.
\end{eqnarray}

Looking at the equations \eqref{5.1}-\eqref{5.3},
one can check the conformal invariance conditions of the dual background.
Hence, the vanishing of the beta-function equations up to two-loop order for the metric \eqref{4.17} and $\tilde B$-field \eqref{4.18} together
with the dilaton field ${\tilde \Phi} = c_{_0}-\log r$
reduce to the following polynomials
\begin{eqnarray}
{\beta_{_{rr}}^G}&=& \alpha' \Big(\frac{2}{k r^3 (r-1)}\Big) + {\cal O}(\alpha'^2) =0,\nonumber\\
{\beta_{_{tt}}^G}&=& -3 {\beta_{_{xx}}^G} = \alpha' \Big(\frac{-8(2r-1) (r-1)^2}{k^2 r^5}\Big) + {\cal O}(\alpha'^2) =0,\nonumber\\
{\beta_{_{tx}}^B}&=&  \alpha' \Big(\frac{8 (2r +6q -1) (r-1)^2}{\sqrt{3} k^2 r^5}\Big) + {\cal O}(\alpha'^2) =0,\nonumber\\
{\beta^\Phi}&=&2 \Lambda +\alpha' \Big(\frac{96 q (r-1) -8 (r^4+2r^2 -2r+1)}{k^2 r^4}\Big) + {\cal O}(\alpha'^2) =0. \label{5.7}
\end{eqnarray}
The above results render the $\alpha'$ expansion is uncontrollable. Therefore, the dual background fails to satisfy the beta-function equations
which indicates that the corresponding $\sigma$-model is not Weyl invariant, i.e. does not define a critical string theory
in the usual sense.


\section{Conclusions}

We have reviewed aspects of PL T-duality in the presence of spectator fields.
Here we have found a non-trivial and interesting example of PL T-dual $\sigma$-models which
helps in the intent of providing a general classification of 6-dimensional geometries describing supergravity backgrounds, also useful for the AdS/CFT correspondence.
We have derived the $AdS_3 \times S^3$ background from PL T-duality on the semi-Abelian double $(A_2 \otimes 2A_1 , 4A_1)$, plus some spectator fields.
In this way, we were able to find a dual pair for this background and
determined its structure including the horizon and singularity.
The components of dual metric \eqref{4.17} were ill been defined at the regions $r=0$ and $r=1$.
By calculating the scalar curvature and the Kretschmann scalar corresponding to the dual metric, we
concluded that $r=0$ was a true singularity, while $r=1$ was nothing but a single horizon.
Accordingly, the non-Abelian T-duality transformation has been related a solution with no horizon and no
curvature singularity to a solution with a single horizon and a curvature singularity.
Then, we investigated the classical integrability of the $AdS_3 \times S^3$ background and its dual pair, in such a way that
we found their corresponding Lax pairs depending on some spectral
parameters.
We should mention that the study of the integrability of the T-dual $\sigma$-models carried out here could be of
interest to string theory in its quest for integrable string backgrounds.
Finally, in order to guarantee UV finiteness at
quantum level, the vanishing of the one-loop beta-functions for both models was imposed.
As seen, because of the non-zero trace of the structure constants corresponding to the
semi-Abelian double $({\A}_2 \oplus 2{\A}_1 , 4{\A}_1)$, the dual dilaton field obtained in
\eqref{5.5} did not follow the dilaton transformation \eqref{c.16.2}.
However, to check the conformal invariance conditions of the T-dual $\sigma$-models, we have carried out a computer assisted study
and thus concluded that the original
$\sigma$-model including the $AdS_3 \times S^3$ background can be remained conformal even up to two-loop order, while the dual background
given by relations \eqref{4.17} and \eqref{4.18} remained conformal only up to one-loop order.
However, we don't know at the moment whether the resulting dual
background has other meaningful physical interpretation.

Another possible direction of further investigation is to consider
the non-Abelian T-dualization of the $AdS_3 \times S^3 \times T^4$ and $AdS_3 \times S^3 \times S^3 \times S^1$ backgrounds
which have smaller set of isometric coordinates, hence,
needs to choose convenient spectator-dependent matrices similar to what we did in the case of the  $AdS_3 \times S^3$ of the present work.
To embed the 6-dimensional background $AdS_3 \times S^3 $ into 10 dimensions one may introduce the flat metric on the
four-torus
\begin{eqnarray}
{ds}_{_{T^4}}^2=dx^i~dx^i,
\end{eqnarray}
where i = 7,...,10. Here, the T-dual $\sigma$-models can be
also constructed on the semi-Abelian double $({\A}_2 \oplus 2{\A}_1 , 4{\A}_1)$
provided that the four-torus contribution is chosen diagonally in $F_{ij}$ of equation \eqref{4.5}.
We intend to address some of these problems in the future.

\subsection*{Acknowledgements}

The author is greatly indebted to the anonymous referee for the constructive
comments to improve the presentation of this work.

\subsection*{Conflict of Interest}

The author declares no conflict of interest.


\subsection*{Data Availability Statement}

No data was used for the research described in the article.

\end{document}